# L-RCM: a method to detect connected components in undirected graphs by using the Laplacian matrix and the RCM algorithm

Francisco Pedroche*     Miguel Rebollo†     Carlos Carrascosa†

Alberto Palomares†

**Abstract**

In this paper we consider undirected graphs with no loops and multiple edges, consisting of k connected components. In these cases, it is well known that one can find a numbering of the vertices such that the adjacency matrix $A$ is block diagonal with k blocks. This also holds for the (unnormalized) Laplacian matrix $L = D - A$, with $D$ a diagonal matrix with the degrees of the nodes. In this paper we propose to use the Reverse Cuthill-McKee (RCM) algorithm to obtain a block diagonal form of $L$ that reveals the number of connected components of the graph. We present some theoretical results about the irreducibility of the Laplacian matrix ordered by the RCM algorithm. As a practical application we present a very efficient method to detect connected components with a computational cost of $O(m + n)$, being $m$ the number of edges and $n$ the number of nodes. The RCM method is implemented in some comercial packages like MATLAB and Mathematica. We make the computations by using the function *symrcm* of MATLAB. Some numerical results are shown.

**Keywords:** ordering algorithms, RCM, graph partitioning, graph spectra, Laplacian matrix.

## 1 Introduction

Some applications in engineering deal with complex networks that may vary its structure in time. This variation may affect the number of links as well as the number of nodes that form the network. For example, we can find these dynamic networks in fields such as traffic networks, communication networks, biological networks, social networks, etc. see, e.g., [Newman(2010)] for an introduction to the application of complex network analysis to some fields. When the structure of a network varies in time it is possible that some nodes may be isolated from the rest. One can also find that groups of nodes may constitute connected components (i.e., nodes connected among them) but disconnected to the rest of the initial network. The standard method to detect connected components in networks is the *breadth first search*, BFS see, e.g., [Jungnickel(2008)], [Gross & Yellen (2004)]. This method was created to find the shortest distance from a given node. BFS can be implemented in running time of $O(n + m)$, being $n$ the number of nodes and $m$ the number of edges of the network. For sparse networks these means $O(n)$ but for dense networks we can have the case $m = n^2$. BFS visits first nodes that are closer to the initial node and lists all the neighbors. On the contrary, the method called *depth first search* (DFS), or *backtracking*, visits first those nodes that are at a long distance from the initial node going as deep as possible. The standard algorithm for the method DFS is usually referred to the seminal paper of Tarjan [Tarjan, (1972)]. DFS can be implemented with a time complexity of $O(m)$, see [Jungnickel(2008)] for details. Another variant of the BFS algorithm is the Reverse Cuthill-McKee algorithm (RCM). This method is based on the Cuthill-McKee algorithm, that was originally devised to reduce the bandwidth of symmetric adjacency matrices [Cuthill & McKee (1969)]. RCM can be implemented with a time complexity $O(q_{max}m)$ where $q_{max}$ is the maximum degree of any node [George & Liu (1981)]. Note that for sparse matrices we have that RCM works with time complexity $O(n)$. RCM is included in some mathematical packages like MATLAB and Mathematica and its routine can be found elsewhere. Therefore, RCM is a practical tool that can be implemented in any formulation using matrices as the standard tool. RCM operates to a matrix $A$ and returns a permutation vector such that one can construct a symmetric permutation $PAP^T$ that has a smaller bandwidth than the original one. When some connected components are presented the RCM method leads to a block diagonal matrix similar to $A$[1]. From this block diagonal form it is straightforward to obtain the nodes corresponding to each connected component, as we show in section 4. However, the RCM method is not commonly used to detect components. In this paper we analyze how to use the RCM to detect components. In particular, we are interested in the properties of the Laplacian matrix obtained after being processed by the RCM method. We shall show some interesting theoretical results that characterize the irreducibility of the Laplacian matrix associated to an undirected graph.

*Institut de Matemàtica Multidisciplinària, Universitat Politècnica de València. Camí de Vera s/n. 46022 València. Spain. {pedroche@imm.upv.es}. Supported by Spanish DGI grant MTM2010-18674.

†Departament de Sistemes Informàtics i Computació, Universitat Politècnica de València. Camí de Vera s/n. 46022 València. Spain.

The structure of the paper is the following. In section 2 we recall some definitions from graph theory and matrix analysis. In section 3 we describe the RCM algorithm and related work. In section 4 we give some theoretical results about the irreducibility of the Laplacian matrix. In section 5 we detail a method to detect connected components using the RCM method, including computational cost. In section 6 we give some results of the application of the proposed method, and in section 7 we give some conclusions and future developments.

## 2 Definitions and known results

Let $G = (V, E)$ be a graph, with $V = \{v_1, v_2, \ldots, v_n\}$ a non-empty set of $n$ vertices (or nodes) and $E$ a set of $m$ edges. Each edge is defined by the pair $(v_i, v_j)$, where $v_i, v_j \in V$.

---

[1] This fact was known by E. Cuthill and J. McKee [Cuthill & McKee (1969)].

The adjacency matrix of the graph $G$ is $A = (a_{ij}) \in \mathbb{R}^{n \times n}$ such that $a_{ij} = 1$ if there is an edge connecting nodes $v_i$ and $v_j$, and 0, otherwise. We consider simple graphs, i.e., undirected graphs without loops (i.e. $(v_i, v_j) = (v_i, v_j)$ and $(v_i, v_i) \notin E$) and without multiple edges (i.e. there's only one edge, if any, from vertex $v_i$ to vertex $v_j$). Therefore $A$ is a symmetric real matrix with zeros in its diagonal. The degree $d_i$ of a node $i$ is the number of its adjacent edges, i.e., $d_i = \sum_{j=1}^{n} a_{ij}$. The (unnormalized) Laplacian matrix of the graph is defined as $L = D - A$ where $D$ is the diagonal matrix $D = diag(d_1, d_2, \ldots, d_n)$. Some other definitions of the *Laplacian* matrix can be found in the literature; a good review of definitions can be found in [Boley *et al* (2011)]. A subgraph $G'$ of $G$ is a graph such that $G' = (E', V')$ with $V' \subseteq V$ and $E' \subseteq E$. A path is a sequence of nodes with an edge connecting every two consecutive nodes. A connected graph is a graph with a path connecting any pair of nodes; otherwise, the graph is said to be disconnected. Let $C_i \subseteq V$ be a set of nodes of a graph. We call $\pi_k = \{C_1, C_2, \ldots, C_k\}$ a partition of $G(V, E)$ when $V = \bigcup_{i=1}^{k} C_i$, and $\bigcap_{i=1}^{k} C_i = \emptyset$. A *connected component* of a graph $G = (V, E)$ is a connected subgraph $G_i(C_i, E_i)$ such that no other node in $V$ can be added to $C_i$ while preserving the property of being connected; i.e., a connected component is a maximal connected subgraph. We are interested in partitions $\pi_k$ of a disconected graph $G(V, E)$ such that each subgraph $G_i(C_i, E_i)$ is a connected component.

We recall that a permutation matrix $P$ is just the identity matrix with its rows re-ordered. Permutation matrices are orthogonal, i.e., $P^T = P^{-1}$. A matrix $A \in \mathbb{R}^{n \times n}$, with $n \geq 2$, is said to be reducible[2] if there is a permutation matrix $P$ of order $n$ and there is some integer $r$ with $1 \leq r \leq n-1$ such that

$$P^T A P = \begin{bmatrix} B & C \\ 0 & D \end{bmatrix}$$

where $B \in \mathbb{R}^{r \times r}$, $C \in \mathbb{R}^{r \times (n-r)}$, and $0 \in \mathbb{R}^{(n-r) \times r}$ is a zero matrix. A matrix is said to be irreducible if it is not reducible. It is known (see, e.g., [Horn & Johnson (1999)]) that the adjacency matrix $A$ of a directed graph is irreducible if and only if the associated graph $G$ is strongly connected. For an undirected graph irreducibility implies connectivity. Note that the Laplacian $L = D - A$ is irreducible if and only if $A$ is irreducible. In the following we recall some properties of $L$.

The Laplacian matrix is positive semidefinite: $x^T L x \geq 0$, for all $x \in \mathbb{R}^{nx1}$, or equivalently, $L$ is symmetric and have all eigenvalues nonnegative. Since $L\mathbf{e}_n = \mathbf{0}$, with $\mathbf{e}_n$ the vector of all ones, $L$ has 0 as an eigenvalue, and therefore is a singular matrix. Since $L$ is real symmetric, is orthogonally diagonalizable, and therefore the rank of $L$ is the number of nonzero eigenvalues of $L$. The eigenvalues of $L$ can be ordered as

$$0 = \lambda_1 \geq \lambda_2 \geq \ldots \geq \lambda_{n-1} \geq \lambda_n \tag{1}$$

It is known that $\lambda_2 = 0$ if and only if $G$ is disconnected. In fact, $\lambda_2$ is called the *algebraic connectivity* of $G$. It is known (see, e.g., [Mohar (1991)], [Abreu(2007)], [Molitierno (2008)], [Nascimento & De Carvalho (2011)]) that the number of connected components of the graph is given by the the algebraic multiplicity of zero as an eigenvalue of the Laplacian matrix;

---

[2]A matrix $A \in \mathbb{R}^{1 \times 1}$ is said to be reducible if $A = 0$.

i.e., the number of components of $G$ is the algebraic multiplicity of $\lambda_2$. Note therefore that the spectrum of $L$ characterizes the connectivity properties of the associated graph $G$.

Note that $L$ is irreducible if and only if $G$ is connected. Therefore $L$ is irreducible if and only if $\lambda_2 \neq 0$. Therefore, since the eigenvalues of $L$ are ordered as in (1) $L$ is irreducible if and only if $rank(L) = n-1$. It is known (see, e.g., [Bolten *et al* (2011)]) that $L$ is a singular M-matrix, i.e., $L$ can be written as $L = sI - B$ with $B \geq 0$ and $s$ the spectral radius of $B$

We recall the following theorem due to Geiringer (see [Varga (2000)] for details) that we shall use later.

**Theorem 2.1** *Let $A = [a_{ij}]$ be an $n \times n$ complex matrix, $n \geq 2$ and let $\mathcal{N} = \{1, 2, \ldots, n\}$. Then it holds that $A$ is irreducible if and only if, for every two disjoint nonempty subsets $S$ and $T$ of $\mathcal{N}$ with $S \cup T = \mathcal{N}$, there exists an element $a_{ij} \neq 0$ of $A$ with $i \in S$, $j \in T$.*

Note that for $n = 1$ we have the trivial case $A = 0 \in \mathbb{R}$ and $L = 0 \in \mathbb{R}$, which is a reducible matrix.

The Laplacian matrix has been used extensively in *spectral clustering* (i.e, the technique of dividing a graph in connected components based on the eigenvectors) see [Fortunato(2010)], [Schaeffer (2007)]. It is known that the inspection of the eigenvectors associated with $\lambda_2$ can lead to a partition of $G$ in connected components. The seminal paper in this field is from Fiedler, see [Fiedler(1973)], [Fiedler(1975)]. The algorithm of the spectral bisection is given in [Pothen *et al* (1989)]. A good explanation can be found in [Newman(2010)]. This technique can be also used to reduce the envelope-size (the sum of the row-widths), see [Del Corso & Romani(2001)], [Kumfert & Pothen (1997)].

**Example 2.1** *Consider the toy graph in Figure 1. A simple computation shows that the spectrum of $L$ is $\lambda_1 = 0, \lambda_2 = 0, \lambda_3 = 2, \lambda_4 = 2$. The algebraic multiplicity of $\lambda = 0$ is two and therefore the graph has two connected components. An easy computation shows that the eigenspace of $\lambda = 0$ is spanned by the vectors*

$$\mathbf{v}_1 = \begin{bmatrix} 1 \\ 0 \\ 1 \\ 0 \end{bmatrix}, \mathbf{v}_2 = \begin{bmatrix} 0 \\ 1 \\ 0 \\ 1 \end{bmatrix}$$

*Note that the indices of the nonzero components of $\mathbf{v}_1$ and $\mathbf{v}_2$ give the labels of the nodes corresponding to each component of $G$.*

# 3 The reverse Cuthill-McKee (RCM) algorithm

The Cuthill-McKee algorithm (CM) was introduced in [Cuthill & McKee (1969)] where the authors presented an scheme to numbering the nodes of a network in such a way that the corresponding adjacency matrix has a narrow bandwidth. They restricted to symmetric positive definite matrices although this condition is not a limitation of their scheme; at

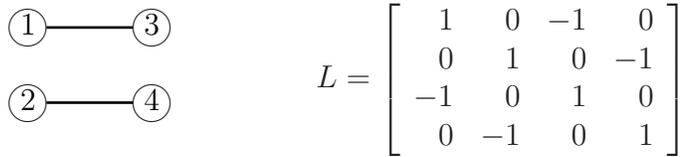

Figure 1: Toy graph with two connected components and its corresponding Laplacian matrix.

that time they were motivated by systems of equations that resulted in these special kind of matrices, see[George & Liu (1981)] for details. To our knowledge the only limitation of their scheme is the symmetry of the adjacency matrix. The main motivation to obtain a narrow bandwidth was to reduce computer storage issues and calculation time. They explain that their objective was to determine a permutation matrix $P$ such that the nonzero elements of $PAP^T$ will cluster about the main diagonal. Therefore, they focused on bandwidth minimization. Their strategy was to search for only few permutations $P$ and then they choose the better one. In most cases (e.g., when the matrix can be transformed to band diagonal with no zero coefficients in the band) their scheme obtained the optimum numbering. The numbering obtained by their scheme corresponds, in graph theory, to the generation of a spanning tree (i.e., a subgraph of G which is a tree containing all the nodes of G, when G is connected). The algorithm selects an starting node and then visits all the neighbors in a level-by-level fashion, as in BFS. They remarked that as a result of their scheme one can easily check whether a matrix is reducible or not. This is the main property that we analyze in this paper. In a later study, [George (1971)], the Reverse Cuthill-McKee algorithm (RCM) was introduced. RCM consists in the CM numbering but in reverse form. This simply procedure worked better than the original CM algorithm in some experimental studies while the bandwidth remained the same. A theoretical comparison of CM and RCM algorithms was given in [Liu & Sherman (1976)] where it is shown that the reverse ordering is always as good as the original one in terms of storage and number of operations. RCM determines a starting node and looks for all the neighbors of this starting node. In a second step, the algorithm looks for the neighbors of these neighbors, and so on. The RCM algorithm can be outlined as follows:

---

**Algorithm 1** RCM (Reverse Cuthill-McKee)

1: select a starting node (e.g. a node with the lowest degree)
2: add $s$ to the queue $Q$
3: **for all** $i = 2$ to $n$ **do**
4:   **for all** $v$ neighbor of $u_i$  **do**
5:     add $v$ to $Q$ in increasing order of degree
6:   **end for**
7: **end for**
8: reverse the order of $Q$

Once obtained the order given by the RCM method, the construction of the matrix $P = (p_{ij})$ is straightforward: in row $i$, $p_{ij} = 1$ with $i$ the new node label and $j$ denotes the original one [Cuthill & McKee (1969)].

Since the introduction of the RCM method some new methods to reduce the bandwidth of a symmetric matrix has been introduced. We remark here that this problem is, in general NP-complete [Mohar & Juvan (1993)]. These new methods can obtain better results in terms of bandwidth reduction but may be more expensive, in computational cost, than the RCM method (see, e.g., [Kumfert & Pothen (1997)]). Since we are interested in using the RCM as a method to detect components we shall not analyze other methods that may optimize the reduction of the bandwidth. See, for example, [Kumfert & Pothen (1997)] for methods like the RCM and methods that use hybrid techniques (spectral plus ordering) to reduce bandwidth. A review of some methods, including a spectral method that uses RCM as a preprocessor can be found in [Del Corso & Romani(2001)]. For a comparison of reorderings, including RCM, in experiments with nonsymmetric matrices associated with the numerical solution of partial differential equations, see [Benzi *el at* (1999)]. For a general view of reordering methods in linear systems, see [Saad (2003)]. A method to extend RCM to unsymmetric matrices is shown in [Reid & Scott (2006)]. RCM has been also used as an inspection tool for graph visualization [Mueller *et al* (2007)].

In most mathematical packages there exists a function that provides the reordering produced by the RCM algorithm. For example, in MATLAB there exists the function *symrcm*. The expression $v = symrcm(A)$ computes an RCM ordering of $A$. This vector $v$ is a permutation such that, using the syntax of MATLAB, $A(v,v)$ tends to have its nonzero entries closer to the diagonal. In MATLAB the algorithm first finds a certain vertex (a *pseudoperipheral* vertex) of the graph associated to the matrix $A$. Therefore the algorithm generates a level structure by breadth-first search and orders the vertices by decreasing distance from this initial vertex [Gilbert *et al* (1992)]. MATLAB claims that the implementation is based closely on the SPARSPAK implementation described in [George & Liu (1981)].

We are interested in applying the RCM algorithm to the Laplacian matrix $L$. To that end, we denote $\widehat{L} = PLP^T$ where $P$ is the permutation matrix obtained with the RCM algorithm applied to $L$. We say that $\widehat{L}$ is a Laplacian matrix of the graph reordered by RCM.

It is known (see, e.g., [Reid & Scott (2006)], [Saad (2003)]) that $\widehat{L}$ is block tridiagonal, i.e.:

$$\widehat{L} = PLP^T = \begin{bmatrix} \widehat{L}_{11} & \widehat{L}_{12} & \cdots & 0 & 0 \\ \widehat{L}_{21} & \widehat{L}_{22} & \widehat{L}_{23} & 0 & 0 \\ 0 & \widehat{L}_{32} & \widehat{L}_{33} & \widehat{L}_{34} & 0 \\ 0 & 0 & \ddots & \ddots & \ddots \\ 0 & 0 & 0 & \widehat{L}_{r,r-1} & \widehat{L}_{r,r} \end{bmatrix} \qquad (2)$$

**Example 3.1** *Given*

$$L = \begin{bmatrix} 1 & 0 & 0 & -1 \\ 0 & 1 & -1 & 0 \\ 0 & -1 & 2 & -1 \\ -1 & 0 & -1 & 2 \end{bmatrix}$$

*The RCM algorithm, computed by MATLAB, gives a permutation vector $\mathbf{v}^T = [2, 3, 4, 1]$. That means, for example, that the old node 2 is now the new node 1, that is $p_{12} = 1$. Therefore the permutation matrix is*

$$P = \begin{bmatrix} 0 & 1 & 0 & 0 \\ 0 & 0 & 1 & 0 \\ 0 & 0 & 0 & 1 \\ 1 & 0 & 0 & 0 \end{bmatrix}$$

*and $\widehat{L}$ results the tridiagonal matrix*

$$\widehat{L} = PLP^T = \begin{bmatrix} 1 & -1 & 0 & 0 \\ -1 & 2 & -1 & 0 \\ 0 & -1 & 2 & -1 \\ 0 & 0 & -1 & 1 \end{bmatrix}$$

## 4 Theoretical Results

In the results shown in this section we use the following notation:

$L \in \mathbb{R}^{n \times n}$ is the Laplacian matrix of an undirected graph $G$.

$\mathcal{N} = \{1, 2, \ldots, n\}$.

$\widehat{L}$ is a matrix derived from the RCM algorithm applied to $L$, that is, $\widehat{L} = PLP^T$.

$\widehat{l}_{ij}$ denotes the $ij$-element of $\widehat{L}$

$\mathbf{1}_i \in \mathbb{R}^{i \times 1}$ a vector column of all ones.

$\mathbf{e}_i \in \mathbb{R}^{n \times 1}$ the vector with ones in the first $i$ entries and zero elsewhere. That is, $\mathbf{e}_i^T = [\mathbf{1}_i^T, \mathbf{0}^T]$. For example, taking $n = 4$, $\mathbf{e}_2^T = [1, 1, 0, 0]$.

We recall that the permutation matrices are orthogonal matrices: $P^T = P^{-1}$. As a consequence, $L$ and $\widehat{L}$ are similar matrices and therefore they have the same spectrum. In detail, $\widehat{L}\mathbf{u} = \lambda \mathbf{u} \to PLP^T\mathbf{u} = \lambda\mathbf{u} \to LP^T\mathbf{u} = \lambda P^T\mathbf{u}$ therefore $\lambda$ is an eigenvalue of $L$ with eigenvector $P^T\mathbf{u}$. Since $L$ and $\widehat{L}$ have the same spectrum we conclude that they have the same rank. Therefore we have that $L$ is irreducible if and only if $\widehat{L}$ is irreducible. It is well known (see, e.g., [Fortunato(2010)]) that for a graph $G$ of $k$ connected components the corresponding Laplacian matrix $L$ can be written as a block diagonal matrix. This fact can be derived from Theorem 2.1. What we want to remark in the following result is that the RCM method gives a permutation matrix $P$ such that $\widehat{L}$ is block diagonal. We give the proof for completeness and also to set the notation that we use later.

**Lemma 4.1** $L$ is reducible if and only if $\widehat{L}$ is a block diagonal matrix of the form:

$$\widehat{L} = \begin{bmatrix} \widehat{L}_{11} & & & \\ & \widehat{L}_{22} & & \\ & & \ddots & \\ & & & \widehat{L}_{kk} \end{bmatrix} \quad (3)$$

with $\widehat{L}_{kk}$ irreducible matrices or $1 \times 1$ zero matrices.

**Proof.** If $L$ is reducible we know that $G$ has $k \geq 2$ connected components with $k$ the multiplicity of $\lambda = 0$. We know that the RCM method detects these $k$ connected components. In fact, the RCM method packs the components. These connected components appear as irreducible blocks $\widehat{L}_{ii}, i = 1, 2, \ldots k$ in $\widehat{L}$; From Theorem 2.1 it is clear that $\widehat{L}$ must be block diagonal. In case of nodes isolated we have $1 \times 1$ null blocks in the diagonal. To prove the theorem in the other direction, note that it is clear that there exists a permutation matrix $P$, given by the RCM method, such that $\widehat{L} = PLP^T$ is block diagonal and therefore, by definition, $L$ is reducible. ∎

In the following theorem we give a first characterization of the irreducibility of $\widehat{L}$.

**Theorem 4.1** Let $n > 1$. $\widehat{L}$ is irreducible if and only if $\widehat{L}\,\mathbf{e}_i \neq \mathbf{0}$ for $i = 1, 2, \ldots, (n-1)$.

**Proof.** Let $\widehat{L}$ be irreducible. Let us assume that there exists $\mathbf{e}_i$, with $i < n$, such that $\widehat{L}\mathbf{e}_i = \mathbf{0}$. Therefore $\mathbf{e}_i$ is an eigenvector for $\lambda = 0$. But since $\mathbf{e}_n$ is also an eigenvector for $\lambda = 0$ we conclude that the eigenspace for $\lambda = 0$ has a dimension greater than 1 and therefore, we conclude that the algebraic multiplicity of $\lambda = 0$ is greater than 1 and then the number of connected components in $\widehat{L}$ is greater than 1 and then $\widehat{L}$ is reducible, which is a contradiction. To prove the theorem in the opposite direction, let us assume that $\widehat{L}$ is reducible. Therefore, $L$ is also reducible and from Lemma 4.1 we have that $\widehat{L}$ must be block diagonal. Therefore there exists $i < n$ (with $i$ the size of the first block, $\widehat{L}_{ii}$) such that $\widehat{L}\,\mathbf{e}_i = \mathbf{0}$, which is a contradiction. Therefore $\widehat{L}$ is irreducible. ∎

We now present a new result from which a first method to detect components can be derived. We prevent the reader that the main result of this paper is another method that we present at the end of this section and we describe in detail in section 5.

We have seen in Lemma 4.1 that when $L$ is reducible then $\widehat{L}$ is block diagonal, with irreducible blocks $\widehat{L}_{ii}$. In the practice (for example, using MATLAB) the RCM method give us a permutation vector to construct $\widehat{L}$. The following result can be used to detect the sizes of the blocks $\widehat{L}_{ii}$ when $\widehat{L}$ is known.

**Theorem 4.2** Let $n > 1$. Let $p \in \mathcal{N}$. If there exists $p < n$ such that:

$$\widehat{L}\,\mathbf{e}_i = \begin{cases} \neq \mathbf{0} & \text{for} \quad i = 1, 2, \ldots, p-1 \\ \mathbf{0} & \text{for} \quad i = p \end{cases} \quad (4)$$

then $\widehat{L}$ is a block diagonal matrix of the form

$$\widehat{L} = \begin{bmatrix} \widehat{L}_{11} & \\ 8 & \widehat{L}_{22} \end{bmatrix} \quad (5)$$

with $\widehat{L}_{11}$ a matrix of size $p \times p$ such that $\widehat{L}_{11}$ is irreducible if $p > 1$ and a zero matrix if $p = 1$.

**Proof.** From Theorem 4.1 we have that $\widehat{L}$ is reducible. Therefore $L$ is reducible and from Lemma 4.1 we have that $\widehat{L}$ is a block diagonal matrix of the form (3). We also know that $\widehat{L}_{11}$ is an irreducible matrix. What we have to prove here is that the size of $\widehat{L}_{11}$ is $p \times p$.

Let us assume that the size of $\widehat{L}_{11}$ is $q \times q$, with $q > 1$, and we study two cases:

Case 1. Suppose that $q < p$. Since $\widehat{L}_{11}$ is itself a Laplacian matrix we have that $\mathbf{1}_q$ is an eigenvector associated to $\lambda = 0$, and therefore $\widehat{L}_{11}\mathbf{1}_q = \mathbf{0}$ and then $\widehat{L}\mathbf{e}_q = \mathbf{0}$ and this is a contradiction with the hypothesis (9).

Case 2. Suppose that $q > p$. By hiphotesis we have $\widehat{L}\,\mathbf{e}_p = \mathbf{0}$, that is

$$\widehat{L}\mathbf{e}_p = \begin{bmatrix} \widehat{L}_{11} & 0 \\ 0 & \widehat{L}_{22} \end{bmatrix} \begin{bmatrix} \mathbf{1}_p \\ \mathbf{0}_{n-p} \end{bmatrix} = \begin{bmatrix} \mathbf{0}_p \\ \mathbf{0}_{n-p} \end{bmatrix} \qquad (6)$$

Let us make a partition of $\widehat{L}_{11}$ in the form

$$\widehat{L}_{11} = \begin{bmatrix} M & T \\ T^T & N \end{bmatrix}$$

with $M$ of size $p \times p$. Therefore, from (6) we have

$$\widehat{L}\mathbf{e}_p = \begin{bmatrix} M & T & 0 \\ T^T & N & 0 \\ 0 & 0 & \widehat{L}_{22} \end{bmatrix} \begin{bmatrix} \mathbf{1}_p \\ \mathbf{0}_{q-p} \\ \mathbf{0}_{n-q} \end{bmatrix} = \begin{bmatrix} \mathbf{0}_p \\ \mathbf{0}_{q-p} \\ \mathbf{0}_{n-q} \end{bmatrix} \qquad (7)$$

Note that $T^T\mathbf{1}_p = \mathbf{0}_{q-p}$ and since $T$ has only negative elements this implies $T = 0$ and therefore

$$\widehat{L}_{11} = \begin{bmatrix} M & 0 \\ 0 & N \end{bmatrix}$$

and then $\widehat{L}_{11}$ is reducible, which is a contradiction.

Therefore we must have $q = p$, and since $\widehat{L}_{11}$ is itself a Laplacian matrix we have $\widehat{L}\,\mathbf{e}_p = \mathbf{0}$ in agreement with (9).

The case $q = 1$ is trivial to prove. Therefore the proof is done. ∎

Note that with a recursive application of this theorem we can compute the block diagonal form of $\widehat{L}$ and thus the disconnected components of the graph $G$. Note that we have to apply the RCM method only once, since the submatrices $\widehat{L}_{22}$, $\widehat{L}_{33}$, etc., are results of the RCM method and we can apply Theorem 4.2 to each one.

**Example 4.1** *Given*

$$L = \begin{bmatrix} 1 & 0 & -1 & 0 \\ 0 & 1 & 0 & -1 \\ -1 & 0 & 1 & 0 \\ 0 & -1 & 0 & 1 \end{bmatrix}$$

*Applying the RCM algorithm to L we obtain that there exists a permutation matrix P such that*

$$\widehat{L} = PLP^T = \begin{bmatrix} 1 & -1 & 0 & 0 \\ -1 & 1 & 0 & 0 \\ 0 & 0 & 1 & -1 \\ 0 & 0 & -1 & 1 \end{bmatrix}$$

*and this matrix verifies conditions (9) with $p = 2$. According to Theorem 4.2 we have that $\widehat{L}$ is a block diagonal matrix with irreducible blocks.*

*An easy computation shows that the eigenspace of $\lambda = 0$ is spanned by the vectors*

$$\mathbf{w}_1 = \begin{bmatrix} 1 \\ 1 \\ 0 \\ 0 \end{bmatrix}, \mathbf{w}_2 = \begin{bmatrix} 0 \\ 0 \\ 1 \\ 1 \end{bmatrix}$$

*Here we see that Theorem 4.2 is inspired in the search of an eigenvector as $\mathbf{w}_1$. Note, however, that the use of Theorem 4.2 avoids the explicit computation of the eigenspace of $\lambda = 0$ to detect components.*

Theorem 4.2 can be used to detect components when $\widehat{L}$ is known.

Nevertheless, making some numerical experiments we find an improved technique to detect components with less computational cost. This technique, which is the main goal of this paper, is inspired in the above theoretical results and based on the following properties and results.

## Properties of the RCM ordering

**Definition**[root of a component]. Given a connected component ordered by RCM we call the *root* of the component the node with the maximum index. We denote by $r$ the index of the root.

Note that this root corresponds to the starting node of the RCM algorithm shown in section 3. Note that by reversing the numbering the root has the maximum index.

**Basic properties**: The following properties follow from the definition of root.

- A connected component has only one root.

- For each node $i$ (different from the root) of the connected component we have $i < r$ with $r$ the index of the root

- If the connected component has $n$ nodes, and the ordering begins in 1, then $r = n$.

- From each node $i$ there is a path to the root $r$ (since the component is connected, and thus irreducible).

**Lemma 4.2** *Given a connected component ordered by RCM, each node $i$, other than the root $r$, is adjacent to a node $i + k \leq r$ for some $k \in \{1, 2, \ldots\}$.*

**Proof.** This is a consequence of the ordered given by the RCM algorithm. Starting from the root (initially node number 1) the algorithm explores the neighbors and goes numbering 2, 3, etc., until it reaches the end of the component having $n$ nodes. Then the RCM algorithm numbers in revers order, i.e., the last visited node is numbered node 1, the following is numbered 2, etc., until the algorithm reaches the root, which has the number $n$. Therefore each node (except the root) is a neighbor of a node with a higher index. ∎

**Example 4.2** *Let us consider the graphs shown in Figure 2. Graph a) is a possible RCM ordering since each node is connected to a node with higher index, except node 3, which is the root. In graph b) we see that node 2 is not connected to a node with higher index, and 2 it is not the root, therefore this is not an ordering given by RCM.*

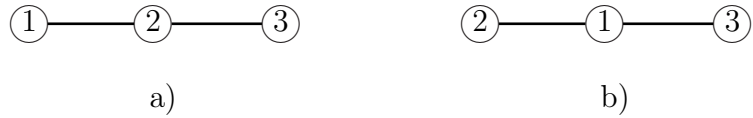

Figure 2: a) is an RCM ordering, while b) it is not.

As a consequence of Lemma 4.2 we have the following.

**Lemma 4.3** *Given a connected component ordered by RCM let $r$ be the root. Let $\widehat{L}$ be the Laplacian of this connected component. Let $s_i(\widehat{L})$ be the sum of the $i$-th row of $\widehat{L}$ up to the diagonal, that is:*

$$s_i(\widehat{L}) = \sum_{j=1}^{i} \widehat{l}_{ij}, \quad \forall i \in \mathcal{N}. \tag{8}$$

*Then $s_i(\widehat{L}) = 0$ if and only if $i = r$.*

**Proof.** We denote $\widehat{L} = PLP^T$, $\widehat{A} = PAP^T$ and $\widehat{D} = PDP^T$ Since $L = D - A$ we have $\widehat{L} = \widehat{D} - \widehat{A}$ with $\widehat{l}_{ii} = \widehat{d}_{ii} = \sum_{j=1}^{n} \widehat{a}_{ij}$.

Therefore

$$s_i(\widehat{L}) = \sum_{j=1}^{i} \widehat{l}_{ij} = \sum_{j=1}^{i-1} \widehat{l}_{ij} + \widehat{l}_{ii} = \sum_{j=1}^{i-1} \widehat{l}_{ij} + \sum_{j=1}^{n} \widehat{a}_{ij} = -\sum_{j=1}^{i-1} \widehat{a}_{ij} + \sum_{j=1}^{n} \widehat{a}_{ij} = \sum_{j>i}^{n} \widehat{a}_{ij}$$

In the last equality we have used that $\widehat{a}_{ii} = 0$ for all $i$. From Lemma 4.2 we have that each node $i$ of the component, except the root, verifies $\widehat{a}_{i,i+k} = 1$ for some $k \in \{1, 2, \ldots\}$ and therefore for all these nodes we have $s_i(\widehat{L}) \neq 0$. The only node that verifies the equality $s_i(\widehat{L}) = 0$ is the root, since it has no links to nodes with higher indices. to it. ∎

Let us now see the properties of the RCM in a graph with more than one connected component.

**Lemma 4.4** Let $\widehat{L}$ be the Laplacian of a graph with $k$ connected components ordered by RCM. Let $r_i$, $i = 1, 2, \ldots k$, be the root of each connected component. Then

1. $s_i(\widehat{L}) = 0$ if and only if $i = r_i$.

2. The nodes of a component $i$ do not have a path to nodes with indices higher than $r_i$.

**Proof.** From Lemma 4.1 we have that $\widehat{L}$ is a block diagonal matrix. In each block $\widehat{L}_{ii}$ we have one component. Therefore Part 1 is a consequence of Lemma 4.3 and Part 2 is a consequence of the fact that the nodes of a component do not belong to other component (if so, the two components would be one component, which is a contradiction). Regarding the numbering, note that from Lemma 4.2 the nodes of component $i$ are described by entries $\widehat{a}_{i,i+k}$ with $i + k \leq r_i$. ∎

**Theorem 4.3** Let $\widehat{L} \in \mathbb{R}^{n \times n}$ be the Laplacian of a graph ordered by RCM. Let $s_i(\widehat{L})$ be given by (8). Then $\widehat{L}$ is irreducible if and only if $s_i(\widehat{L}) = 0$ only occurs for $i = n$.

**Proof.** If $\widehat{L}$ is irreducible then there is only one component and from Lemma 4.3 the proof follows. In the opposite direction, since $s_n(\widehat{L}) = 0$ from Lemma 4.3 we have that $n$ is the root and from the basic properties we know that any node can follow a path to reach the root. Therefore the graph associated to matrix $\widehat{L}$ is connected and therefore $\widehat{L}$ is irreducible. ∎

**Theorem 4.4** Let $\widehat{L} \in \mathbb{R}^{n \times n}$ be the Laplacian of a graph ordered by RCM. Let $s_i(\widehat{L})$ be given by (8). Let $n \geq 2$. Let $p \in \mathcal{N}$. If there exists $p < n$ such that:

$$s_i(\widehat{L}) = \begin{cases} \neq 0 & \text{for} \quad i = 1, 2, \ldots, p-1 \\ 0 & \text{for} \quad i = p \end{cases} \tag{9}$$

then $\widehat{L}$ is a block diagonal matrix of the form

$$\widehat{L} = \begin{bmatrix} \widehat{L}_{11} & \\ & \widehat{L}_{22} \end{bmatrix} \tag{10}$$

with $\widehat{L}_{11}$ a matrix of size $p \times p$ such that $\widehat{L}_{11}$ is irreducible if $p > 1$ and a zero matrix if $p = 1$.

**Proof.** Since $s_i(\widehat{L}) = 0$ for $i \neq n$ we have, from Theorem 4.3 that $\widehat{L}$ is reducible. From Lemma 4.1 we have that $\widehat{L}$ is block diagonal with more than one block. Since Since $s_p(\widehat{L}) = 0$ we have, from Lemma 4.4 that node $p$ is a root of a component. The nodes of this component must have indices lower than $p$. Therefore, the $p$ nodes of this component are the nodes described by $\widehat{L}_{11}$. In the particular case that $p = 1$ then $\widehat{L}_{11}$ has only one element, which is zero: this means that node node 1 is an isolated node. It is a component of a single element which is the root. ∎

By a recursive application of Theorem 4.4 one can derive a fast practical method to detect components. We show the details in the next section.

**Example 4.3** *Given the matrices $L$ and $\widehat{L}$ of example 4.1 we have that $s_i(\widehat{L}) = 0$ for $i = 2$ and $i = 4$. This means that there are two roots and therefore $\widehat{L}$ has two components. $\widehat{L}$ is a block diagonal with two blocks: one formed by the nodes $1, 2$ and the other by the nodes $3, 4$ (with the numbering given by RCM, which is the used in $\widehat{L}$).*

## 5 A practical method to detect connected components

The main goal of this paper is to present a fast method to detect components. The method is based on Theorem 4.4. The method can be written, in MATLAB notation, in the following form:

---
**Algorithm 2** L-RCM Algorithm
---
1: $L = sparse(diag(sum(A)) - A);$
2: $rcm = symrcm(L);$
3: $Lp = L(rcm, rcm);$
4: $s = sum(tril(Lp), 2);$
5: $cut = find(s == 0);$
---

In the first line we construct the Laplacian matrix from the adjacency matrix. In the experiments we use sparse matrices. The number of nonzero entries of $A$ are $nnz = 2m$, being $m$ the number of edges. Since $A$ is symmetric it only suffices to handle $m$ entries. In sparse matrices, $m$ is much less than $n^2$. To sum the $2m$ elements of $A$ we require a computational cost of $O(2m)$ and therefore to construct $L$ we have a cost of order $O(3m)$.

The second line of the algorithm computes the RCM vector, that is, the ordering given by the RCM algorithm as computed by MATLAB. We have noted, in section 3, that MATLAB follows a procedure given in [George & Liu (1981)]. Therefore, as we have noted in the Introduction, we can assume a time complexity of $O(q_{max}m)$, where $q_{max}$ is the maximum degree of any node.

The third line computes the Laplacian matrix of the graph ordered by the RCM method. This matrix is what we have called $\widehat{L}$ in previous sections. Obviously, it would have a very high cost to compute $\widehat{L}$ as the explicit matrix product $PLP^T$, with $P$ the permutation matrix derived from the RCM. According to [MATLAB R2012a Documentation (2012)] the cost of reordering a matrix, as in line 3, is proportional to $nnz$ of that matrix. Therefore let us denote by $O(\gamma\, nnz(\widehat{L}))$ the cost of computing $Lp$, with $\gamma$ some parameter that depends on the implementation in MATLAB. Since $nnz(\widehat{L}) = nnz(A) + n = 2m + n$ we have that the cost of the third line of the algorithm is of the order $O(\gamma(2m + n))$.

In line 4 the algorithm computes the sums $s_i(\widehat{L})$ defined in equation (8). To compute these sums in MATLAB we take the lower triangular part of $\widehat{L}$ and we sum each row. Here we have $m$ entries of $A$ to sum plus the diagonal of $L$ that counts $n$ entries. Therefore this operation has a cost of $O(m + n)$.

Finally, in line 5 we find the indices of the sums that verify $s_i(\widehat{L}) = 0$. Let us assume that this operation costs $O(n)$. The indices of these sums, the entries of the vector *cut*, give the location of the roots of the components, according to Theorem 4.4, i.e., vector *cut* gives the indices where there exists a gap in the blocks.

Therefore the total cost of the algorithm is of the order:

$$O([q_{max} + 2\gamma + 4]m + (2 + \gamma)n) \approx O(m + n)$$

The outputs of the method are the permutation vector *rcm* that defines the new ordering and the *cut* vector that identifies the blocks over the order defined by *rcm*.

In the following example we show how the method works. In section 6 we explore the cost of compute $Lp$ with MATLAB. We also give the results concerning the relative part of the time consumed by each part of the algorithm.

**Example 5.1** *Let us consider the graph shown in Figure 3.*

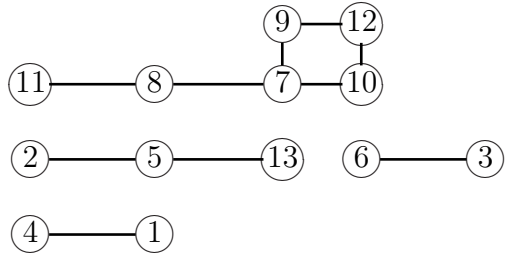

Figure 3: Graph with $n = 13$ and four connected components

*Matrix $L$ is given by*

$$L = \begin{bmatrix} 1 & 0 & 0 & -1 & 0 & 0 & 0 & 0 & 0 & 0 & 0 & 0 & 0 \\ 0 & 1 & 0 & 0 & -1 & 0 & 0 & 0 & 0 & 0 & 0 & 0 & 0 \\ 0 & 0 & 1 & 0 & 0 & -1 & 0 & 0 & 0 & 0 & 0 & 0 & 0 \\ -1 & 0 & 0 & 1 & 0 & 0 & 0 & 0 & 0 & 0 & 0 & 0 & 0 \\ 0 & -1 & 0 & 0 & 2 & 0 & 0 & 0 & 0 & 0 & 0 & 0 & -1 \\ 0 & 0 & -1 & 0 & 0 & 1 & 0 & 0 & 0 & 0 & 0 & 0 & 0 \\ 0 & 0 & 0 & 0 & 0 & 0 & 3 & -1 & -1 & -1 & 0 & 0 & 0 \\ 0 & 0 & 0 & 0 & 0 & 0 & -1 & 2 & 0 & 0 & -1 & 0 & 0 \\ 0 & 0 & 0 & 0 & 0 & 0 & -1 & 0 & 2 & 0 & 0 & -1 & 0 \\ 0 & 0 & 0 & 0 & 0 & 0 & -1 & 0 & 0 & 2 & 0 & -1 & 0 \\ 0 & 0 & 0 & 0 & 0 & 0 & 0 & -1 & 0 & 0 & 1 & 0 & 0 \\ 0 & 0 & 0 & 0 & 0 & 0 & 0 & 0 & -1 & -1 & 0 & 2 & 0 \\ 0 & 0 & 0 & 0 & -1 & 0 & 0 & 0 & 0 & 0 & 0 & 0 & 1 \end{bmatrix}$$

*and vector rcm is*

$$rcm = [4, 1, 13, 5, 2, 6, 3, 11, 8, 7, 10, 9, 12]$$

*Matrix $\widehat{L}$ is given by*

$$\widehat{L} = \begin{bmatrix} 1 & -1 & 0 & 0 & 0 & 0 & 0 & 0 & 0 & 0 & 0 & 0 & 0 \\ -1 & 1 & 0 & 0 & 0 & 0 & 0 & 0 & 0 & 0 & 0 & 0 & 0 \\ 0 & 0 & 1 & -1 & 0 & 0 & 0 & 0 & 0 & 0 & 0 & 0 & 0 \\ 0 & 0 & -1 & 2 & -1 & 0 & 0 & 0 & 0 & 0 & 0 & 0 & 0 \\ 0 & 0 & 0 & -1 & 1 & 0 & 0 & 0 & 0 & 0 & 0 & 0 & 0 \\ 0 & 0 & 0 & 0 & 0 & 1 & -1 & 0 & 0 & 0 & 0 & 0 & 0 \\ 0 & 0 & 0 & 0 & 0 & -1 & 1 & 0 & 0 & 0 & 0 & 0 & 0 \\ 0 & 0 & 0 & 0 & 0 & 0 & 0 & 1 & -1 & 0 & 0 & 0 & 0 \\ 0 & 0 & 0 & 0 & 0 & 0 & 0 & -1 & 2 & -1 & 0 & 0 & 0 \\ 0 & 0 & 0 & 0 & 0 & 0 & 0 & 0 & -1 & 3 & -1 & -1 & 0 \\ 0 & 0 & 0 & 0 & 0 & 0 & 0 & 0 & 0 & -1 & 2 & 0 & -1 \\ 0 & 0 & 0 & 0 & 0 & 0 & 0 & 0 & 0 & -1 & 0 & 2 & -1 \\ 0 & 0 & 0 & 0 & 0 & 0 & 0 & 0 & 0 & 0 & -1 & -1 & 2 \end{bmatrix}$$

and vector cut is

$$cut = [2, 5, 7, 13]^T$$

From vectors rcm and cut we have the components (with the original labelling): $\{4, 1\}$, $\{13, 5, 2\}$, $\{6, 3\}$ and $\{11, 8, 7, 10, 9, 12\}$. In figure 4 we show the graph with the ordered given by the vector rcm.

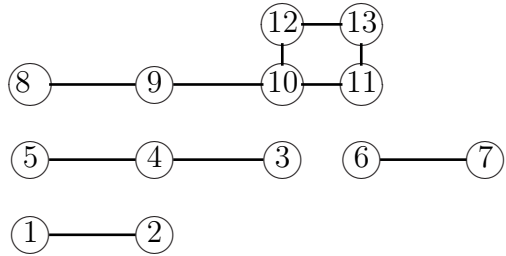

Figure 4: Graph with $n = 13$, ordered with RCM.

# 6 Numerical results

## 6.1 Constant number of nodes and blocks of variable size

In these experiments we create disconnected graphs with a variable number of blocks, while keeping $n$ constant. We investigate the time consumed in running each part of Algorithm 2.

As an input matrix to the algorithm we create, with random generation, a network of $n = 2^{16}$ nodes with $k$ connected components of the same size. We give a random numbering to this graph. For each $k$ we run the algorithm 10 times and we compute the mean cost. We prove with $k = 2^p$, with $p = 5, 6, \ldots, 13$. The results, normalized using the total cost of the algorithm for the case of two blocks ($p = 1$, not shown in the figure), are summarized in Figure 5. In this figure we call the following parts of algorithm 2: Red square ($\square$): $L$, the part corresponding to line 1. Blue bullet ($\bullet$): $RCM$, the part corresponding to line 2. Green Diamond ($\diamond$): $Lp$, the part corresponding to line 3. Black Triangle ($\triangle$): sf, (*sum and find*) the part corresponding to lines 4 and 5.

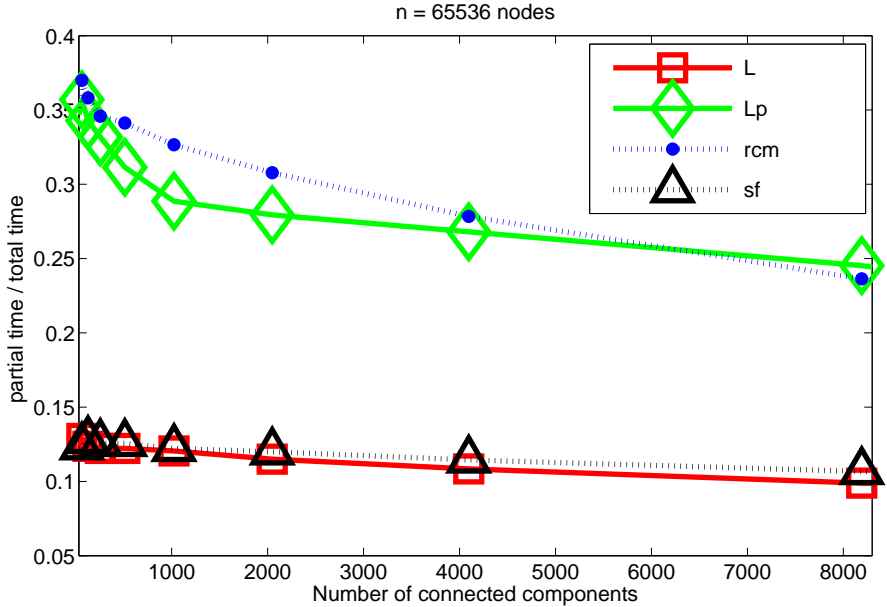

Figure 5: Experiments with $n$ approximately constant. Note that the cost of computing $Lp$ (that is, $\widehat{L}$) is of the same order of computing the RCM. The parts *sum and find* and the computation of $L$ are of the same order.

For example, for $k = 12$ we have a network with $2^{16}$ nodes formed by $2^{12} = 4096$ connected components of $2^4$ nodes each. For $k = 13$ we have a network with $2^{16}$ nodes formed by $2^{13} = 8192$ connected components of $2^3$ nodes each.

In the experiments the number of nodes varies slightly from $2^{16} = 65536$ to $63125$ in the run with $k = 13$ blocks. This is because, in generating randomly the blocks, the number of edges is not constant and, in breaking up the blocks, the total number of nodes decreases. In the experiments we have $q_{max}$ ranging from 15 corresponding to $k = 5$ to $q_{max} = 7$ corresponding to $k = 13$ blocks. Regarding $m$ it varies from $m \approx 2n$ to $m \approx 1.4n$. As a consequence we have that the RCM as the Lp parts of the algorithm remains more o less constant with a tendency to decrease due to these decreasing values of $m$ and $q_{max}$.

The values of the parts $L$ and $sf$ also remains constant with a slight decreasing due to the decreasing in the values of $m$.

Regarding the relative costs we note, for example, that for $k = 7$ blocks we have $m \approx 2n$ and $q_{max} = 15$ from which the cost of RCM is $O(q_{max}m) \approx O(15 * 2n) = O(30n)$. While the cost of the $Lp$ parts is $O(\gamma(2m + n) \approx O(\gamma 5n)$. Since the relative costs appear to be similar in the experiments we assume that in this case $\gamma \approx 6$.

With the details of the case for $k = 13$ blocks we have $m \approx 1.4n$ and $q_{max} = 7$ from which the cost of RCM is $O(q_{max}m) \approx O(7*1.4n) \approx O(10n)$. While the cost of the $Lp$ part is $O(\gamma(2m+n)) \approx O(\gamma 3.8n)$. Again, since the relative costs appear to be similar in the experiments, we assume that in this case $\gamma \approx 2.6$.

As we see, the factor $\gamma$ that affects the computation of the $Lp$ part with MATLAB is a parameter of the same order of $q_{max}$, in these experiments.

Regarding the relative cost of the parts $L$ and $sf$ we find, that for $k = 7$ blocks we have $m \approx 2n$ then the cost of $L$ is $O(6n)$ and the cost of the part of $sf$ is $O(4n)$, which means that they are of the same order as the experiments show.

With $k = 13$ blocks we have $m \approx 1.4n$, and then the cost of $L$ is $O(4.2n)$ and the cost of the part of $sf$ is $O(3.4n)$, which also means that they are of the same order as the experiments show.

## 6.2 Increasing number of nodes with two connected components

In these experiments we generate two blocks and then merge them into a graph of $n$ nodes. The algorithm detects the two blocks. We register the total time of the algorithm. The results are shown in figure 6 and figure 7.

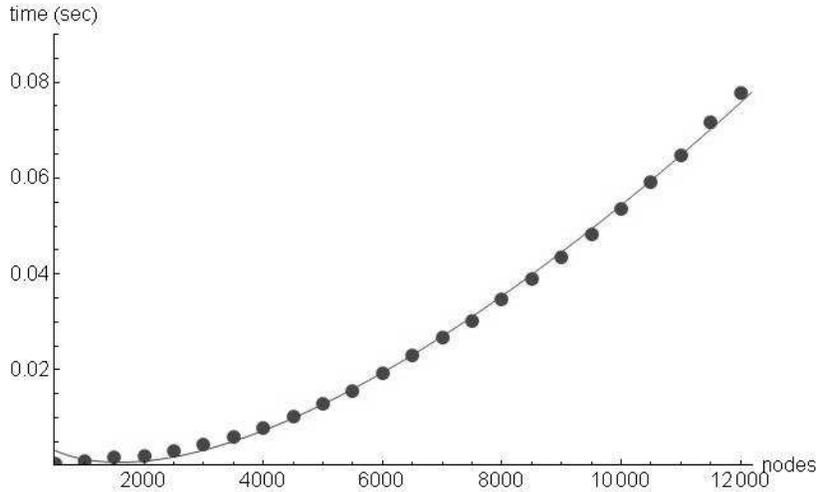

Figure 6: Experiments with two blocks and $n$ increasing. Sparsity: $nnz(A)/n^2 = 0.005$.

From these figures we see that the total time of the algorithm results nearly proportional to $n$, since the fitted curve is of the form $an^{1.1} + bn + c$ in both cases. In particular, in figure 6, the fitted curve is of the form $an^{1.1} + bn + c$. With $a = 0.008$, $b = 5 \times 10^{-5}$, $c = 2 \times 10^{-5}$. We have $q_{max} = 92$ for $n = 12000$.

In figure 7, the fitted curve is of the form $an^{1.1} + bn + c$. With $a = 0.08$, $b = 6 \times 10^{-4}$, $c = 2 \times 10^{-4}$. We have $q_{max} = 667$ for $n = 12000$.

We also note the influence of the quotient $m/n$ in the resulting total time. These trends are in accordance with the analysis of the computational cost of the algorithm.

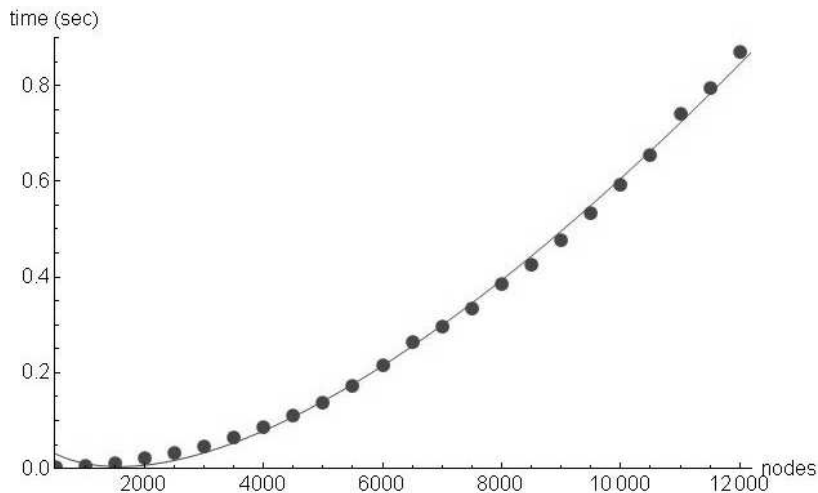

Figure 7: Experiments with two blocks and $n$ increasing. Sparsity: $nnz(A)/n^2 = 0.05$.

# 7 Conclusions

In this paper we have shown a practical method to detect connected components in undirected graphs. The method, called L-RCM (see algorithm 2 in section 5), is based in two ingredients. On one hand, we use the Laplacian matrix to describe the network. On the other hand, we use the RCM algorithm to symmetrically permute the Laplacian matrix. This permutation of the Laplacian matrix reveals the number of the connected components in the graph; to show this, we give some theoretical results concerning the irreducibility of the permuted Laplacian matrix. The main result, Theorem 4.3, allows to derive a practical method to detect components with a complexity of $O(m+n)$ being $m$ the number of edges and $n$ the number of nodes. The method is algebraic and it is exact. We give some examples using MATLAB to show how the experimental computational cost follows the theoretical predictions.

# Acknowledgments

This work is supported by Spanish DGI grant MTM2010-18674, Consolider Ingenio CSD2007-00022, PROMETEO 2008/051, OVAMAH TIN2009-13839-C03-01, and PAID-06-11-2084.